\documentclass[iop]{emulateapj}

\usepackage{color}
\usepackage{graphicx}
\usepackage{amsmath}
\usepackage[colorlinks,linkcolor=blue,anchorcolor=blue,citecolor=blue]{hyperref}

\newcommand\dx{{\rm d}}
\newcommand\arcsinh{{\rm arcsinh}}
\newcommand\sinn{{\rm sinn}}
\newcommand\jcap{JCAP}
\renewcommand\prl{PhRvL}
\renewcommand\prd{PhRvD}
\renewcommand\apjl{ApJL}

\shorttitle{The Cosmological Redshift Relation}
\shortauthors{Tian}

\begin{document}
\title{The relation between cosmological redshift and scale factor for photons}
\author{Shuxun Tian}
\affil{Department of Astronomy, Beijing Normal University, Beijing 100875, China; \emph{tshuxun@mail.bnu.edu.cn}\\
  Department of Physics, Wuhan University, Wuhan 430072, China}

\begin{abstract}
  The cosmological constant problem has become one of the most important ones in modern cosmology. In this paper, we try to construct a model that can avoid the cosmological constant problem and have the potential to explain the apparent late-time accelerating expansion of the universe in both luminosity distance and angular diameter distance measurement channels. In our model, the core is to modify the relation between cosmological redshift and scale factor for photons. We point out three ways to test our hypothesis: the supernova time dilation; the gravitational waves and its electromagnetic counterparts emitted by the binary neutron star systems; and the Sandage--Loeb effect. All of this method is feasible now or in the near future.
\end{abstract}

\section{Introduction}\label{sec:01}
Through the observations of Type Ia supernovae (SNe Ia), \citet{Riess1998} and \citet{Perlmutter1999} found that the universe is accelerating, and the $\Lambda$ cold dark matter ($\Lambda$CDM) model could be well fitted to the observed data. This fact is also confirmed by subsequent observations about SNe Ia \citep{Suzuki2012}, baryon acoustic oscillations \citep[BAO;][]{Eisenstein2005,Anderson2014,Ata2017}, and cosmic microwave background \citep[CMB;][]{Bennett2013,Planck2016}, etc. Besides the cosmological constant, the scalar field can also play a role as dark energy, such as quintessence \citep{Caldwell1998,Carroll1998,Zlatev1999} or k-essence \citep{Armendariz2000,Armendariz2001} field. In addition, the modified gravity theory can also be used to explain the acceleration of the universe \citep{Clifton2012,Joyce2015}. These models can be well fitted to the observations, but theoretically there are various problems. For example, most models lack physical motivation. Another important issue is the cosmological constant problem, which includes a coincidence problem and a fine-tuning problem. The dynamic dark energy models can generally be used to alleviate the coincidence problem \citep{Amendola2010}, but this does not help with the cosmological constant problem. The fine-turning problem is exacerbated in the interacting dark energy models \citep{Marsh2017}.

In this paper, we propose a new idea to explain the apparent acceleration of the late-time universe, which is totally different with the dark energy models or the modified gravity theories. First, almost all of the previous astronomical observations are based on electromagnetic waves (EWs). With the detection of gravitational waves \citep[GWs;][]{Abbott2016a,Abbott2016b}, we are about to enter the era of GW astronomy. There are many discussions on the application of GWs in cosmology \citep[e.g.][]{Schutz1986,Cutler2009,Nishizawa2011,Nishizawa2012,Camera2013}, most of which discuss the precision of the cosmological parameter constraints. Here, we ask whether the same cosmological results can be obtained from GW and EW observations alone. If not, what can we learn from it? For the present observations, GWs can be well described by general relativity \citep{Abbott2016c} without the need for any quantum effects, but photons (EWs) are different. The discussion about photons led the development of early quantum mechanics. Therefore, it is more reasonable to attribute the possible cosmological difference between GW and EW observations to some unknown quantum properties of photons, and this may reveal clues to current acceleration scenarios.

For now, we are mainly concerned with the possibility of explaining the late-time acceleration by phenomenally modifying some properties of photons in the Friedmann-Lema\^{i}tre-Robertson-Walker (FLRW) metric. The possible unknown property of photons has been widely discussed in cosmologies. Historically, the tired light hypothesis has been used to explain the cosmological redshift of photons \citep{Zwikcy1929}. In this model, the universe is static, but photons lose energy during transit. With the discovery of CMB \citep{Penzias1965}, the static cosmology model has gradually faded. More recently, the tired light hypothesis has also been directly ruled out by the supernovae time dilation data \citep{Goldhaber2001,Blondin2008}. For the acceleration of the late universe, \citet{Csaki2002} discussed whether this acceleration is due to the possible interaction between the photon and a pseudoscalar field. In their model, photons can convert into some unseen particles during propagation over cosmological distances, which makes the observed supernova brightness darker. An important feature of this model is that the angular diameter distance is not affected by the new interaction, which makes the model able to be excluded by the measurement of the CMB acoustic peak and BAO signals \citep{Song2006}. Besides the possible interaction, intergalactic dust may also make the supernova brightness darker \citep{Aguirre1999a,Aguirre1999b}. All of these models only affect the luminosity distance without affecting the angular diameter distance. In this paper, we start from the modification of the relationship between cosmological redshift and scale factor for photons, and expect to explain the observations with a flat matter-dominated universe. As we will see, both the luminosity distance and the angular diameter distance will be affected by the modification. This is an essential difference between our model and models with dimming of light.

This paper is organized as follows. In section \ref{sec:02}, we introduce the hypothesis for photons. In section \ref{sec:03}, we discuss three possible astronomical tests for the hypothesis. In section \ref{sec:04}, we derive the cosmological distances in the matter-dominated universe. Our conclusions will be summarized in section \ref{sec:05}.

\section{Hypothesis}\label{sec:02}
The universe is described by the FLRW metric
\begin{equation}
  \dx s^2=-c^2\dx t^2+a^2(t)\left[\frac{\dx r^2}{1-Kr^2}+r^2\dx\Omega^2\right].
\end{equation}
For the applicability of general relativity, the motion of a single photon still needs to be described by the null geodesic. If a photon is in the position of $r=r_e$ at cosmic time,  $t_e$, propagates inward along the radial direction, and reaches the coordinate origin at $t_0$, then the equation of motion can be written as
\begin{equation}
  c\dx t=-\frac{a(t)\dx r}{\sqrt{1-Kr^2}}.
\end{equation}
Integration yields
\begin{equation}\label{eq:03}
  \int_{t_e}^{t_0}\frac{c\dx t}{a(t)}=%
  \int_{0}^{r_e}\frac{\dx r}{\sqrt{1-Kr^2}}=%
  \left\{
    \begin{array}{ll}
      \displaystyle\frac{\arcsin{\sqrt{K}r_e}}{\sqrt{K}} & K>0 \\
      \displaystyle r_e & K=0. \\
      \displaystyle\frac{\arcsinh{\sqrt{-K}r_e}}{\sqrt{-K}} & K<0
    \end{array}
  \right.
\end{equation}
Thus,
\begin{equation}
  r_e=\frac{\sinn\left[\sqrt{|K|}\displaystyle\int_{t_e}^{t_0}\frac{c\dx t}{a(t)}\right]}{\sqrt{|K|}},
\end{equation}
where
\begin{equation}
  \sinn(x)=\left\{
  \begin{array}{ll}
    \sin(x) & K>0 \\
    x & K=0. \\
    \sinh(x) & K<0
  \end{array}
  \right.
\end{equation}
Assuming that another photon is in the position of $r=r_e$ at cosmic time $t_e+\delta t_e$, propagates inward along the radial direction, and reaches the coordinate origin at $t_0+\delta t_0$. Combined with equation (\ref{eq:03}), we can obtain
\begin{equation}
  \frac{\delta t_e}{a_e}=\frac{\delta t_0}{a_0},
\end{equation}
where $a_e\equiv a(t_e),\ a_0\equiv a(t_0)$. $\delta t$ can be regarded as the time interval between two macro events. The classical relation between cosmological redshift and scale factor for photons is
\begin{equation}
  1+z=\frac{a_0}{a_e}=\frac{\delta t_0}{\delta t_e}.
\end{equation}
We call this the classical redshift relation hereafter. If we assume the photon frequency is significantly affected by some unknown quantum effects during the propagation on cosmic scales, then the redshift relation can be phenomenologically modified to be
\begin{equation}
  1+f(z)=\frac{a_0}{a_e}=\frac{\delta t_0}{\delta t_e},
\end{equation}
where $f(z)$ satisfies $f(0)=0$, $f'(z)\equiv\dx f/\dx z>0$ when $z\geq0$. We assume that the frequency of photons is affected by some unknown microscopic quantum effects, but the time dilation of two macroscopic events caused by the expansion of the universe will not be affected, i.e. $a_0/a_e=\delta t_0/\delta t_e$ is still applicable. We intend to use the above modification combined with the flat matter-dominated universe to explain the apparent accelerating expansion of the late universe. In this model, because of the modification of the redshift relation for photons, an EW observation at any cosmic time will deviate from the prediction of the classical matter-dominated universe. This provides a very natural way to solve the coincidence problem. This advantage also exists in the previous models with dimming of light. In the dynamic evolution of the universe, we do not need the vacuum energy. Therefore, a mechanism that make quantum vacuum contribute nothing to the gravity \citep[e.g. take the quantum field theory cutoff to infinity in the theory proposed by][]{Wang2017}, could be used to solve the cosmological constant problem \citep[more precisely the ``old" cosmological constant problem; for more discussions, see][]{Weinberg1989}.

\section{Astronomical Test}\label{sec:03}
\begin{figure}
	\includegraphics[width=\columnwidth]{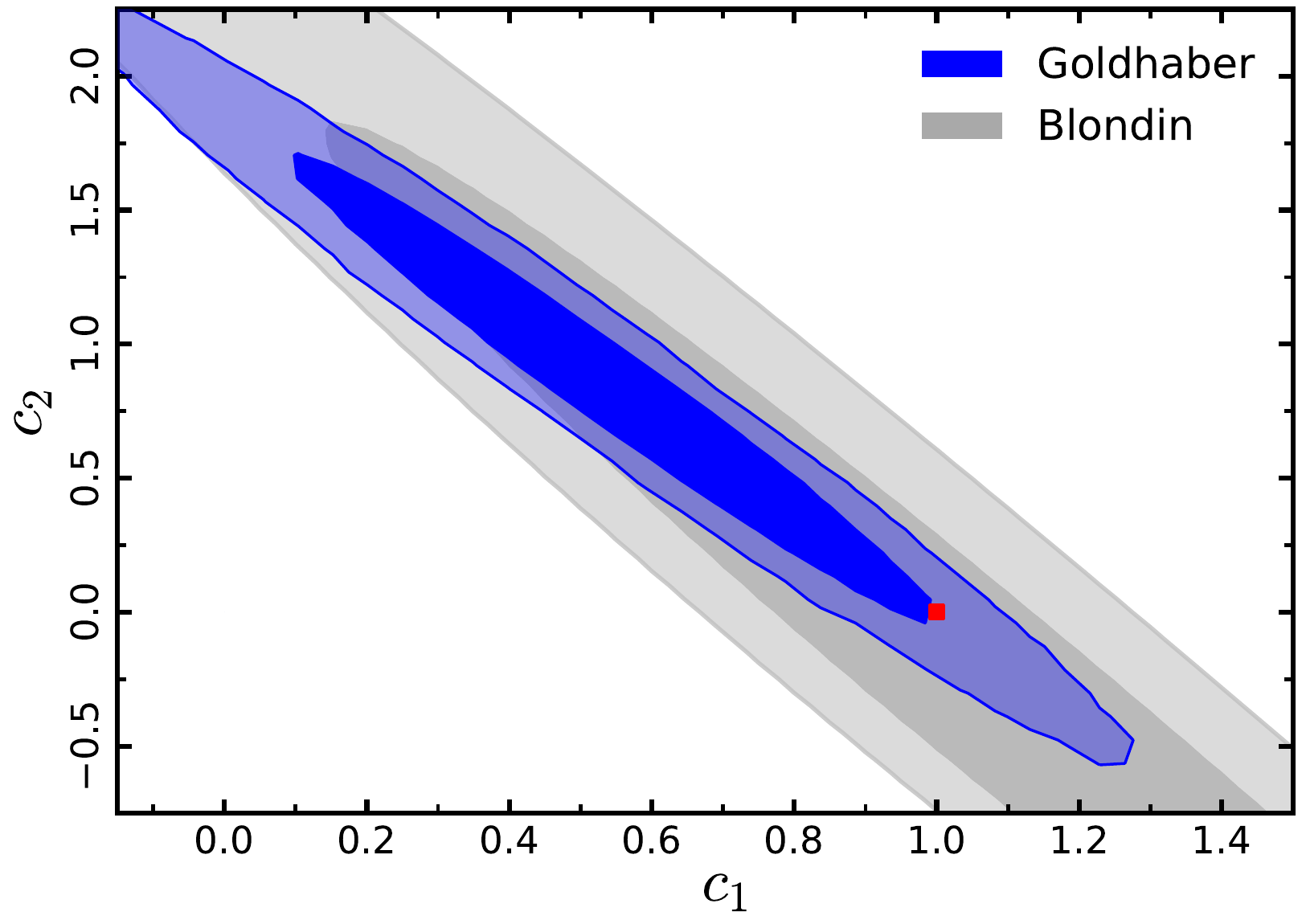}
    \caption{Constraints from the supernovae time dilation data. More specifically, light curve width factor \citep[][blue contours]{Goldhaber2001} and spectra aging rate \citep[][gray contours]{Blondin2008}. The deep and light colors correspond to the confidence intervals of $1\sigma$ and $2\sigma$, respectively. The red square corresponds to the classical redshift relation.}
    \label{fig:01}
\end{figure}

In fact, the above hypothesis can be directly verified by the time dilation data of SNe Ia. \citet{Goldhaber2001} gives 42 time-axis width factors, $w_{\rm ob}$ ($\equiv\delta t_0/\delta t_e$, that denote the stretch of the supernova light curve) using the B-band light curve of SNe Ia (see Table 1 in G01). In this paper, we are mainly concerned with 34 of the total 42 data points (this excludes seven data with comments in Table 1, and SN 1992bi because of its stretch factor, which is too large). As G01 did, we also need to take into account the intrinsic dispersion. We use $w_{\rm th}=1+c_1z+c_2z^2$ to fit the data, then the likelihood \citep{DAgostini2005} can be written as
\begin{align}
  \mathcal{L}(c_1,c_2,\delta_w)\propto&\prod_{i=1}^{34}\frac{1}{\sqrt{\sigma_{w,i}^2+\delta_w^2}}\nonumber\\
  &\times\exp\left[-\frac{(w_{{\rm th},i}-w_{{\rm ob},i})^2}{2(\sigma_{w,i}^2+\delta_w^2)}\right]
\end{align}
where $\delta_w$ is the intrinsic dispersion. We perform a Markov Chain Monte Carlo algorithm {\tt DRAM}\footnote{\url{http://helios.fmi.fi/~lainema/dram/}}\citep{Haario2006} to maximizing the above likelihood, and use the {\tt Python} package {\tt corner}\footnote{\url{http://github.com/dfm/corner.py}} to plot contours. The blue in Fig.~\ref{fig:01} shows the joint distribution of $c_1$ and $c_2$ for this data, and the result indicates that existing data does allow or prefer to modify the classical redshift relation. The constraint on $\delta_w$ is $\delta_w=0.11^{+0.03}_{-0.02}$ at $68\%$ confidence level, which is agreement with the choice in G01. In addition, \citet{Blondin2008} gives 13 aging rate$(\equiv\delta t_e/\delta t_0)$ data points using the spectrum of SNe Ia (see Table 3 in B08). As discussed in B08, to keep the errors of Gaussian, we should use $1/w_{\rm th}$ to fit the data. Note that here we do not have to include the intrinsic dispersion because of the poor data quality. The gray in Fig.~\ref{fig:01} shows the $1\sigma$ and $2\sigma$ confidence intervals constraint from this data. This result also allows modification of the classical redshift relation. In fact the supernova light curve may have a potential evolution with redshift (see comments in B08), which may make corresponding data \citep[G01 or the JLA data;][]{Betoule2014} can not be used to test our hypothesis. In contrast, the aging rate data used in B08 can provide a more correct constraint.

Here, we propose two other ways to test our hypothesis, which is executable in the near future. One is the GW and EW signals produced by the binary neutron star (BNS) systems. \citet{Messenger2012} and \citet{Messenger2014} pointed out that the redshift of the BNS systems can be obtained from the third-generation GW observations alone. As we discussed before, current observation shows that GWs can be well described by general relativity \citep{Abbott2016c} without the need for any quantum effects. Thus, the classical redshift relationship should still available for GWs. So far, the redshift relation of GWs and photons can be written as
\begin{gather}
  1+z_{\rm GW}=\frac{a_0}{a_e}=\frac{\delta t_0}{\delta t_e},\\
  1+f(z_\gamma)=\frac{a_0}{a_e}=\frac{\delta t_0}{\delta t_e},
\end{gather}
respectively. Meanwhile, it is generally believed that the electromagnetic signals can be produced during the BNS mergers \citep[e.g.][]{Paczynski1986,Li1998,Ciolfi2017}, which can provide the redshift information of photons. Because it is the same source, we have $z_{\rm GW}=f(z_\gamma)$, so that we can use the GW and EW signals generated by BNS systems to test our hypothesis. Furthermore, this observation can be used to reconstruct the expression of $f(z)$.

The other method is the Sandage--Loeb effect \citep{Sandage1962,Loeb1998}, which characterizes the change of the redshift for one astronomical source over time. In future observations such as those of the Extremely Large Telescope%
\footnote{\url{http://www.eso.org/public/teles-instr/elt/}\ \url{http://www.iac.es/proyecto/codex/}} or the Square Kilometer Array\footnote{\url{http://skatelescope.org/}}, this effect is observable \citep{Liske2008,Darling2012}. For the classical Sandage--Loeb effect, the redshift drift can be expressed as
\begin{equation}\label{eq:12}
  \dot{z}=(1+z)H_0-H(z),
\end{equation}
where $H_0$ is Hubble constant, and $H(z)$ is the Hubble parameter in the classical cosmological models. If we modify the redshift relation, then the corresponding formula of Sandage--Loeb effect should also be changed. Assuming we receive the electromagnetic signals at $t_0$ and $t_0+\delta t_0$, which is emitted by a source at $t_e$ and $t_e+\delta t_e$, then\footnote{Essentially $f$ may also explicitly depend on $t_0$, but for simplicity we do not consider this situation here.}
\begin{gather}
  1+f[z(t_0)]=\frac{a(t_0)}{a(t_e)},\\
  1+f[z(t_0+\delta t_0)]=\frac{a(t_0+\delta t_0)}{a(t_e+\delta t_e)}
\end{gather}
Combined with $\delta t_e=[a(t_e)/a(t_0)]\cdot\delta t_0$, and $a(t+\delta t)\approx a(t)+\dot{a}(t)\delta t$, we have
\begin{align}
  &f'(z)\cdot\dot{z}(t_0)\cdot\delta t_0=f[z(t_0+\delta t_0)]-f[z(t_0)] \nonumber\\
  &\quad=\frac{a(t_0+\delta t_0)}{a(t_e+\delta t_e)}-\frac{a(t_0)}{a(t_e)}
   =\left[\frac{\dot{a}(t_0)-\dot{a}(t_e)}{a(t_e)}\right]\delta t_0.
\end{align}
This gives the new formula for the Sandage--Loeb effect with the modified redshift relation
\begin{equation}
  \dot{z}=\frac{(1+f)H_0-H(z)}{f'(z)},
\end{equation}
where $H(z)$ is the Hubble parameter under the modified redshift relationship (see equation (\ref{eq:22}) for an example). For the $\Lambda$CDM model, we have a specific $\dot{z}-z$ relation, i.e. equation (\ref{eq:12}). If the observation deviates from the classical relationship, this can be attributed to the modified redshift relation. Of course, this deviation may also be caused by different classical cosmological models. In summary, the Sandage--Loeb effect provides an indirect test of our hypothesis, while the GW and EW signals produced by BNS systems can provide a direct test.

\section{Cosmological Distance}\label{sec:04}
In this section, we derive the luminosity distance and angular diameter distance for optical observations in the matter-dominated universe with the modified redshift relation. Note that the classical redshift relation may be used in current astronomical data processing. The distance obtained here can not directly fit the existing data.

\subsection{Luminosity distance}
The luminosity distance is defined as
\begin{equation}
  D_L\equiv\sqrt{\frac{L}{4\pi\cdot l}},
\end{equation}
where $L$ is the luminosity of the source, and $l$ is the observed flux density. For simplicity, we assume the source emits $n$ photons with a frequency of $\nu_e$ within the time interval $\delta t_e$, then
\begin{equation}
  L=\frac{nh\nu_e}{\delta t_e}
\end{equation}
where $h$ is the Planck constant. There are three factors to influence the observed flux density:
\begin{itemize}
  \item the observed time interval is stretched, i.e. $\delta t_0=\delta t_e\cdot a_0/a_e=\delta t_e\cdot(1+f)$;
  \item $n$ photons are distributed on a large sphere with a coordinate radius of $r_e$, so the physical area of the sphere is $4\pi a_0^2r_e^2$. Note that this area is independent of the curvature $K$ because $\dx r=0$ on the sphere; and
  \item single photon is redshifted with $\nu_0=\nu_e/(1+z)$.
\end{itemize}
Thus,
\begin{align}
  l=\frac{n}{4\pi a_0^2 r_e^2}\cdot\frac{h\nu_0}{\delta t_0}=\frac{L}{4\pi a_0^2 r_e^2}\cdot\frac{1}{(1+z)(1+f)},
\end{align}
which gives the expression of the luminosity distance
\begin{equation}\label{eq:20}
  D_L(z)=\sqrt{\frac{L}{4\pi l}}=\sqrt{(1+z)(1+f)}\cdot a_0r_e.
\end{equation}
If we want the universe to consist only of dust, then the Friedmann equation is
\begin{equation}
  H^2=\frac{8\pi G}{3}\rho-\frac{Kc^2}{a^2}=\frac{8\pi G\rho_0}{3}\frac{a_0^3}{a^3}-\frac{Kc^2}{a^2}.
\end{equation}
We denote $\Omega_{\rm m}=8\pi G\rho_0/3H_0^2$ and $\Omega_K=-Kc^2/a_0^2H_0^2$ hereafter. Then the Friedmann equation becomes
\begin{equation}\label{eq:22}
  E^2(z)\equiv\frac{H^2(z)}{H_0^2}=\Omega_{\rm m}(1+f)^3+\Omega_K(1+f)^2.
\end{equation}
Combined with
\begin{gather}
  \frac{\dx a}{\dx t}=aH,\quad\frac{\dx a}{\dx z}=-f'\frac{a^2}{a_0}=-a_0\frac{f'}{(1+f)^2},
\end{gather}
we have
\begin{equation}
  \frac{\dx t}{\dx z}=\frac{-f'}{(1+f)H(z)}.
\end{equation}
Thus, the luminosity distance can be simplified to
\begin{align}
  &D_L(z)=\sqrt{(1+z)(1+f)}\cdot a_0r_e\nonumber\\
    &\quad=\sqrt{(1+z)(1+f)}\cdot a_0
      \frac{\sinn\left[\sqrt{|K|}\displaystyle\int_{t_e}^{t_0}\frac{c\dx t}{a(t)}\right]}{\sqrt{|K|}}\nonumber\\
    &\quad=\frac{c\sqrt{(1+z)(1+f)}}{H_0}\cdot
      \frac{\sinn\left[\sqrt{|\Omega_K|}\displaystyle\int_{0}^{z}\frac{f'\dx z}{E(z)}\right]}{\sqrt{|\Omega_K|}}
\end{align}
In fact, the above integral can be expressed more concisely. Especially for the flat universe, we have
\begin{align}
  D_L(z)&=\frac{c\sqrt{(1+z)(1+f)}}{H_0}\cdot\int_{0}^{z}\frac{f'\dx z}{(1+f)^{3/2}}\nonumber\\
    &=\frac{2c}{H_0}\sqrt{1+z}\cdot\left(\sqrt{1+f}-1\right)
\end{align}
Appropriate selection of the free function $f(z)$ can be used to fit the observations.

\subsection{Angular diameter distance}
The source at coordinates $r_e$ shines at $t_e$, and is observed at $t_0$ to subtend a small angle $\delta\theta$, then its tangential physical length is $a(t_e)r_e\delta\theta$. Then we can obtain the angular diameter distance
\begin{equation}
  D_A=\frac{a(t_e)r_e\delta\theta}{\delta\theta}=a(t_e)r_e.
\end{equation}
Combined with equation (\ref{eq:20}), we obtain the new distance duality relation
\begin{equation}
  D_L=\sqrt{(1+z)(1+f)^3}D_A.
\end{equation}
There are many works to test the distance duality relation \citep[e.g.][]{Holanda2016,Liao2016}, and no strong evidence of the violation was found. But as we mentioned before, the processing of the relevant data may use the classical redshift relation, so the current results are not enough to rule out our hypothesis. Note that our model may produce the same effect on the luminosity distance as the models with dimming of light, but they lead to different angular diameter distance predictions.

\section{Conclusion}\label{sec:05}
In this paper, we have proposed a new model to explain the late-time acceleration of the universe. The most attractive feature of the model is that there is no cosmological constant problem. In particular, our model provides a very natural way to solve the coincidence problem. We point out three ways to test our hypothesis: the supernova time dilation; the GWs and its electromagnetic counterparts emitted by BNS systems; and the Sandage--Loeb effect. The BNS systems can provide the most direct test even for one event, and the other two methods need to do constraints statistically. In the matter-dominated universe, we derive the luminosity distance and angular diameter distance under the modified redshift relation, which can be probed with future observations.

\ \par
\emph{Note --- }After the publication of this article, we noticed that \citet{Wojtak2016} also proposed a modification of the classical redshift relation, but for different purpose. We recommand readers to read this paper for different discussions on this topic.

\section*{Acknowledgements}
We are grateful to the referee and editor for valuable comments. This work was supported by the National Natural Science Foundation of China under grant No. 11633001.

\end{document}